\documentclass[aps,pra, twocolumn]{revtex4-1}
\usepackage{hyperref} 
\usepackage{epsfig}
\usepackage{amsmath}
\usepackage{subfigure}
\usepackage{graphicx}

\usepackage{color}

\definecolor{mygreen}{rgb}{0,0.5,0} 
\definecolor{myblue}{rgb}{0,0,0.75} 
\definecolor{myyellow}{rgb}{0.87,0.8,0.47} 
\definecolor{mymagenta}{cmyk}{0,1,0,0.12}

\newcommand{\ain}{a_{\rm in}}
\newcommand{\aout}{a_{\rm out}}
\newcommand{\bin}{b_{\rm in}}

\begin{document}

\title{Theory of high gain cavity-enhanced spontaneous parametric down-conversion }

\author{Joanna A. Zieli\'{n}ska$^{1}$}
\email[Corresponding author  ]{joanna.zielinska@icfo.es}
\author{Morgan W. Mitchell$^{1,2}$}
\affiliation{$^{1}$ICFO -- Institut de Ciencies Fotoniques,
Mediterranean Technology Park,
08860 Castelldefels, Barcelona, Spain \\
$^{2}$ICREA -- Instituci\'o Catalana de Recerca i Estudis
Avan\c{c}ats, 08015 Barcelona, Spain}

\newcommand{\multimode}{multimode~}

\begin{abstract}

We compute the output of \multimode cavity-enhanced spontaneous parametric down-conversion (SPDC) for sub-threshold, but otherwise arbitrary, gain. We find analytic Bogoliubov transformations that allow us to calculate arbitrary field correlation functions, including the second-order intensity correlation function $G^{(2)}(T)$.  The results show evidence of increased coherence due to stimulated SPDC.  We extend an earlier model [Lu and Ou, Phys. Rev. A, {\bf 62}, 033804 (2000)] to arbitrary gain and finesse, and show the extension gives accurate results in most scenarios.  The results will allow simple, analytic description of cavity-based nonclassical light sources for quantum networking, quantum-enhanced sensing of atoms and generation of highly non-classical field states.\end{abstract}

\maketitle

\section{Introduction}

Cavity-enhanced spontaneous parametric down-conversion (CESPDC), in which a spontaneous parametric down-conversion (SPDC) process is resonantly enhanced by placing the $\chi^{(2)}$ medium inside an optical cavity, has been used to make highly efficient photon pair sources  \cite{ScholzAPL2007, WolfgrammJOSAB2010} of  interest for quantum networking with atomic quantum memories \cite{PomaricoNJP2009,ChuuAPL2012, PomaricoNJP2012, FeketePRL2013} and atomic quantum metrology \cite{WolfgrammPRL2010, WolfgrammNPhot2013}, applications that require both high spectral brightness and narrow line widths.  SPDC sources in combination with coherent states have been proposed as extremely bright photon pair sources \cite{BeduiniPRL2013}, and as sources of entangled multi-photon states  \cite{AfekS2010, MitchellNJP2014}. 

Many calculations of the fields emitted by CESPDC are based on techniques developed to calculate squeezing in parametric amplifiers \cite{OuPRL1999,lu2000}.  The cavity is described in a modal expansion and quantum reservoir theory \cite{GardinerPRA1985} is used to derive dynamical relationships between cavity, input, and output fields.  When these are solved, the resulting Bogoliubov transformation expresses the output fields as squeezed versions of the input fields \cite{collett1984,GardinerOC1984}.  Using this approach, Lu and Ou \cite{lu2000} computed $G^{(2)}(T)$, the second-order intensity correlation function for type-I CESPDC.  Reflecting experimental conditions of the time, that calculation remained in the low-gain limit and approximated the cavity line-shapes as Lorentzian, as appropriate to high finesse cavities. 

In contemporary applications, there is a trend toward lower-finesse cavities in CESPDC \cite{MehmetOE2011}.  The available single-pass gain has increased, due to periodically poled nonlinear materials and more powerful pump lasers, and lowering the finesse allows higher escape efficiencies at the same system gain level.  At these lower finesses, the ``tails'' of the modes begin to overlap, and mode shapes deviate from the simple Lorentzian. At the same time, higher-gain applications, for example in generation of ``Schr{\"o}dinger kitten''  \cite{NeergaardNielsenPRL2006} states and other highly non-classical time-domain states \cite{AndersenPRA2013,MorinNPhot2014,JeongNPhot2014} by photon subtraction, are also becoming important.  These higher-gain processes necessarily involve stimulated SPDC \cite{Lamas-LinaresN2001}, in which a photon or a pair of photons induces the production of more pairs.  These developments motivate a new calculation of CESPDC fields beyond the low gain, single-longitudinal-mode, and high-finesse approximations.  

Our method is similar to the classic works of Collett and Gardiner \cite{collett1984} and Gardiner and Savage \cite{GardinerOC1984}, in that we use input-output relations for squeezing and cavity in/out-coupling to obtain equations relating input, output, and intra-cavity fields.  In contrast to those works, we avoid quantum reservoir theory by posing the problem directly in the time domain.  As we describe below, narrow-band CESPDC is more naturally and transparently described in this way.  We find difference equations describing the input, output, and cavity fields at consecutive round-trip times.  Eliminating the cavity field from these equations, we find the Bogoliubov transformation expressing the output fields in terms of the input fields.  
To study the time-domain structure, we calculate the second-order intensity correlation function $G^{(2)}(T)$ for a type-I OPO, including arbitrary finesse and gain.
We find an envelope well approximated by a double exponential with a gain-dependent decay constant, multiplied by a comb structure with a period equal to the cavity round trip time. At low gain and high finesse this agrees with the calculation of \cite{lu2000}. At higher gains we find coherence beyond the cavity ring-down time due to stimulated SPDC.

\section{Bogoliubov transformations}
\begin{figure}
  \centering
  \includegraphics[width=7.5cm]{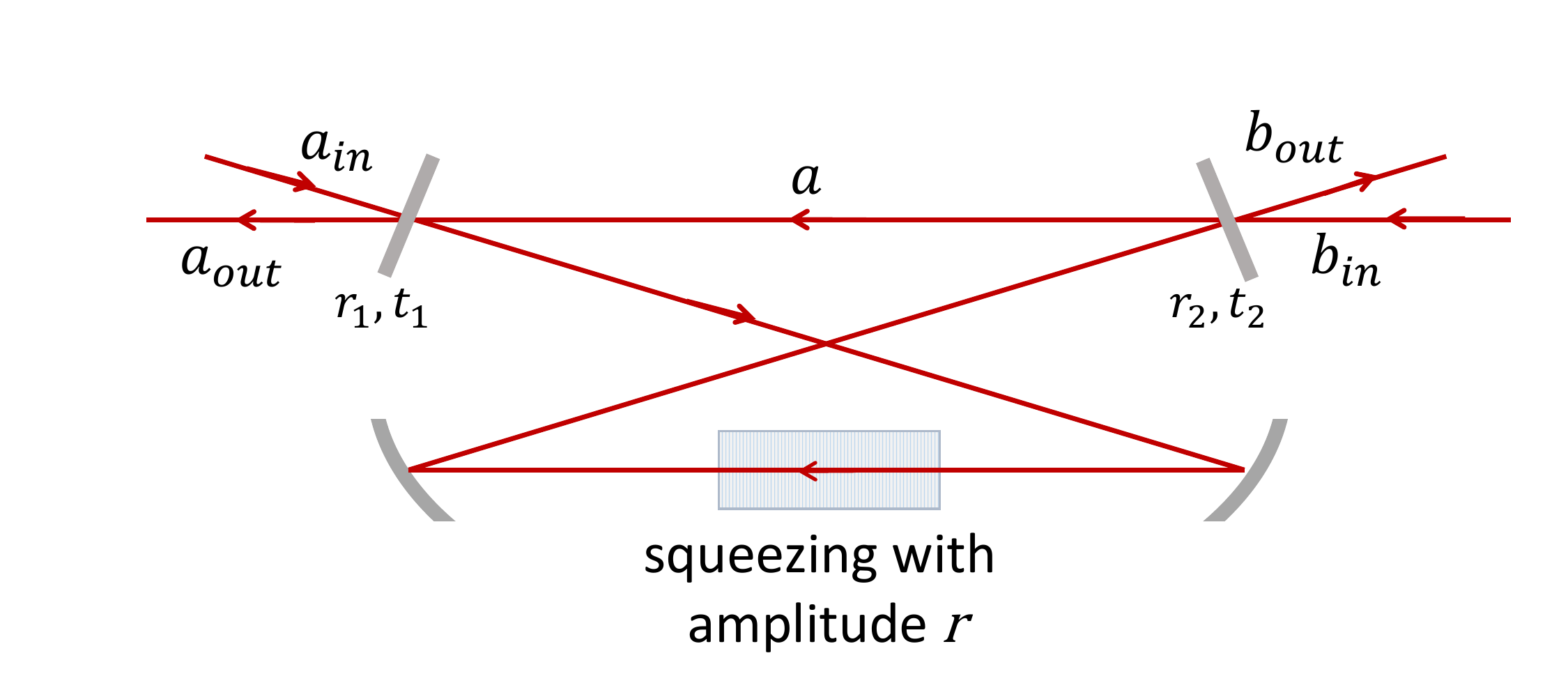}
\caption{(color online) An OPO scheme with input, output and intracavity field operators for double-sided cavity with a nonlinear crystal inside. }
  \label{fig:OPO}
\end{figure}

Let us consider a two-sided ring cavity as in Fig. \ref{fig:OPO} with roundtrip time denoted as $\tau$. We characterize the cavity amplitude transmission and reflection coefficients with real numbers $t_i$ and $r_i$, where a subscript $i=1,2$ indicates the output coupler and another mirror representing the collective cavity losses, respectively. For each of the beamsplitters, there are four numbers describing the input-ouput relation, the transmission from inside the cavity (`c') to the exterior (`e') $t_{i,ce}$, the transmission from the exterior to the interior of the cavity $t_{i,ec}$, the reflection from inside the cavity $r_{i,cc}$ and the reflection from the outside the cavity $r_{i,ee}$. These coefficients are related by energy conservation:  $| t_{i,ce}|^2 + | r_{i,ee}|^2 = | t_{i,ec} |^2 + |r_{i,cc}|^2 = 1$ and $t_{i,ce} r_{i,ec}^* + t_{i,ec} r_{i,cc}^* = 0$. We assume that all $t$ and $r$ coefficients are real, and $t_{i,ec} = t_{i,ce}\equiv t_i $, and $r_{i,cc} = - r_{i,ee}\equiv r_i$.  The intracavity field annihilation operator just before reaching the output coupler is denoted as $a$, while the input fields just before reaching  the cavity are $\ain$ and $\bin$.  We denote the output field just after exiting the cavity as $\aout$.

The field experiences three relevant transformations during a round-trip of the cavity.  Interaction with the output coupler produces 
\begin{eqnarray}
a & \stackrel{{\rm OC}}{\rightarrow} &r_1 a + t_1 \ain,
\end{eqnarray}
where $\ain$ is the input field.  Other losses (here lumped together in a single interaction) produce
\begin{eqnarray}
a & \stackrel{{\rm loss}}{\rightarrow} &r_2 a + t_2 \bin,
\end{eqnarray}
where $\bin$ is a bath mode assumed to be in vacuum.  Finally there is the Bogoliubov transformation due to squeezing on a single pass through the crystal
\begin{eqnarray}
a & \stackrel{{\rm sq}}{\rightarrow} & a \cosh(r) + a^\dagger \sinh(r),
\end{eqnarray}
where $r$ is the squeezing amplitude.

 Applying these three transformations in sequence to $a(t-\tau)$ (understood to be the intra-cavity field at a location immediately before the output coupler), we have
\begin{eqnarray}
AfekS2010a&\rightarrow& r_1 a + t_1 \ain \\
&\rightarrow& r_2(r_1 a + t_1 \ain) + t_2 \bin \\
&\rightarrow& \cosh(r) [ r_2(r_1 a + t_1 \ain) + t_2 \bin] \nonumber \\
& & +\sinh(r) [ r_2(r_1 a^\dagger + t_1 \ain^\dagger) + t_2 \bin^\dagger].
\end{eqnarray}
Considering that a round-trip takes time $\tau$  and the field $a(t)$ depends only on $a(t-\tau)$, which is true if we neglect the dispersion and finite bandwidth of the phase-matching (see below), we have 
\begin{eqnarray}
\label{eq:roundtrip}
a(t)&=&r_1 r_2 \cosh(r)a(t-\tau)+r_1 r_2 \sinh(r)a^\dagger(t-\tau) \nonumber \\ 
& & +t_1 r_2 \cosh(r)\ain(t-\tau) +t_1 r_2 \sinh(r)\ain^\dagger(t-\tau)
\nonumber \\ & & +t_2 \cosh(r)\bin(t-\tau)+t_2 \sinh(r)\bin^\dagger(t-\tau) ~~~
\end{eqnarray}
with the hermitian conjugate:
\begin{eqnarray}
\label{eq:croundtrip}
a^\dagger(t)&=&r_1 r_2 \cosh(r)a^\dagger(t-\tau)+r_1 r_2 \sinh(r)a(t-\tau) \nonumber \\ 
& & +t_1 r_2 \cosh(r)\ain^\dagger(t-\tau) +t_1 r_2 \sinh(r)\ain(t-\tau)
\nonumber \\ & & +t_2 \cosh(r)\bin^\dagger(t-\tau)+t_2 \sinh(r)\bin(t-\tau).
\end{eqnarray}
The output field is given by
\begin{eqnarray}
\label{eq:output}
\aout(t)=-r_1 \ain(t)+t_1 a(t).
\end{eqnarray}

Writing
\begin{eqnarray}
a(t)=\frac{1}{\sqrt{2\pi}}\int_{-\infty}^\infty a(\omega)e^{-i\omega t} d\omega \nonumber\\\nonumber
a^\dagger(t)=\frac{1}{\sqrt{2\pi}}\int_{-\infty}^\infty a^\dagger(\omega)e^{i\omega t} d\omega 
\end{eqnarray}
and solving Eqs. (\ref{eq:roundtrip}),(\ref{eq:croundtrip}), (\ref{eq:output}) for $\aout$, we find the Bogoliubov transformation

\begin{align}
\label{eq:aout}
\aout(\omega)&=A(\omega) \ain(\omega)+B(\omega) \ain^\dagger(-\omega) \nonumber \\
&  +C(\omega) \bin(\omega)+D(\omega) \bin^\dagger(-\omega)
\end{align}
where
\begin{eqnarray}
\label{eq:Adef}
A(\omega)&\equiv& d(\omega) t_1^2r_2[ e^{-i\omega\tau}\cosh(r)-r_1 r_2 ] -r_1 \\
B(\omega)&\equiv& d(\omega) \sinh(r) t_1^2 r_2 e^{-i\omega\tau} \\
C(\omega)&\equiv& d(\omega) t_2 t_1[ e^{-i\omega\tau}\cosh(r)-r_1 r_2 ]  \\
\label{eq:Ddef}
D(\omega)&\equiv& d(\omega) \sinh(r) t_2 t_1 e^{-i\omega\tau} 
\end{eqnarray}
and
\begin{eqnarray}
\label{eq:denominator}
d(\omega)&\equiv& \frac{1}{[e^{-i\omega \tau}-r_1 r_2 \cosh(r)]^2-[r_1 r_2 \sinh(r)]^2 }.
\end{eqnarray}

Eqs. (\ref{eq:aout}) to (\ref{eq:denominator}) constitute a full description of the output of the OPO, in the sense that any correlation function of interest can be calculated by taking expectation values of products of $\aout$ and $\aout^\dagger$.  For example, the degree of quadrature squeezing at a side-band frequency of $\Omega$ can be computed as 
\begin{align}
S(\Omega) &\equiv  \langle [\aout(\Omega) + \aout^\dagger(-\Omega)]^2 \rangle,
\end{align}
where the expectation $\langle \cdot \rangle$ is taken with respect to vacuum in both the $a$ and $b$ modes.  $S(\Omega)$ is simply a polynomial in $A(\Omega)$ to $D(-\Omega)$, so analytical results are available for any gain level. 

We have neglected dispersion in the cavity and the finite phase-matching bandwidth of the crystal. In this case the emission spectrum of the source is not limited by the phase matching profile, and depends only on the cavity parameters. These approximations are justified in typical narrow-band CESPDC scenarios \cite{Anacavity}, in which the phase matching bandwidth is several orders of magnitude larger than the free spectral range (FSR) of the cavity. Introducing a finite phase matching bandwidth would modify the shape of the peaks composing the \multimode $G^{(2)}(T)$, but at a time-scale beyond the resolution of current electronics. As described in \cite{Anacavity} the KTP nonlinear crystal introduces a dispersion of $dn/d\lambda=-0.06~{\rm \mu m}^{-1}$, which over a phase-matching bandwidth of 100 GHz ($\approx$ 0.2 nm) changes the refractive index by  $10^{-5}$, not shifting any of the resonances by more than $10^{-3}~ {\rm FSR}$. In contrast, broad-band CESPDC experiments are typically sensitive to the full output bandwidth of the SPDC process \cite{Jeronimo-MorenoLP2010}, and these approximations would not be justified.  

~

\section{Multimode $G^{(2)}(T)$}
\label{sec:MMG2}
Time-domain correlation measurements on OPOs are an important diagnostic of the spectral content of the output \cite{PomaricoNJP2009,ChuuAPL2012, PomaricoNJP2012, FeketePRL2013}, and are often used to demonstrate the quantum nature of the generated fields \cite{ScholzAPL2007,WolfgrammJOSAB2010}. In this section we compute the intensity correlation function $G^{(2)}(T)$.  As with the degree of squeezing, this can be computed analytically for any sub-threshold gain level and including all modes.  

\newcommand{\vacbra}{\langle{\rm vac}|}
\newcommand{\vacket}{|{\rm vac}\rangle}
\renewcommand{\vacbra}{\langle{0}|}
\renewcommand{\vacket}{|{0}\rangle}

As described above, this correlation function is computed as a normally-ordered expectation value with respect to the vacuum state in both input modes:
\begin{eqnarray}
\label{step1}
G^{(2)}(T)&\equiv&\langle \aout^\dagger(t)\aout^\dagger (t+T)\aout (t+T)\aout(t)\rangle
 \\ 
&=&
\int d^4\omega \,  e^{-i(\omega_2+\omega_3)(t+T)} e^{-i(\omega_1+\omega_4)t} G^{(2)}(\vec{\omega}) \hspace{4mm}
\end{eqnarray}
where $d^4\omega \equiv d\omega_1\, d\omega_2\, d\omega_3\, d\omega_4$ and
\begin{eqnarray}
G^{(2)}(\vec{\omega}) &\equiv& 
\langle \aout^\dagger (-\omega_1)\aout^\dagger (-\omega_2)\aout (\omega_3)\aout(\omega_4)\rangle. \nonumber  
\end{eqnarray}

After the reduction of the operators using the commutation relation $[a(\omega), a^\dagger(\omega')]=\delta(\omega-\omega')$ and knowing that the coefficients $A(\omega)$, $B(\omega)$, $C(\omega)$ and $D(\omega)$ are hermitian functions, e.g. $A(-\omega) = A^*(\omega)$, we find the expression under the Fourier transform
\newcommand{\GG}{GG}

\begin{figure}
  \centering
  \includegraphics[width=6.5cm]{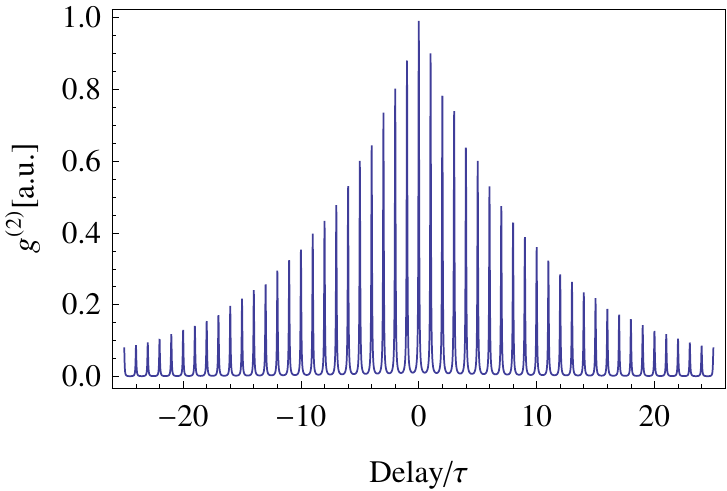}
\caption{(color online) Theoretical $G^{(2)}(T)$ calculated for cavity parameters as for the source presented in \cite{Anacavity} with gain equal to $1\%$ of the OPO threshold. The envelope of the $G^{(2)}(T)$ is calculated from  Eqs. (\ref{step2}), (\ref{eq:GammaFourier}) and (\ref{eq:UpsilonFourier}), and normalized to unity at $T=0$. For the purpose of plotting, the peaks, which in the model are Dirac delta functions, have been replaced with finite-width Lorentzians. }
  \label{theoryplot}
\end{figure}

\begin{small}
\begin{eqnarray}
G^{(2)}(\vec{\omega}) & =& \delta(\omega_1+\omega_2)\delta(\omega_3+\omega_4)\Gamma(\omega_2, -\omega_1)\Gamma(\omega_3, -\omega_4)
 \nonumber  \\ & &  
 +\delta(\omega_2+\omega_3)\delta(\omega_1+\omega_4)\Upsilon(\omega_1, -\omega_4)\Upsilon(\omega_2, -\omega_3)
  \nonumber  \\ & &  
  +\delta(\omega_1+\omega_3)\delta(\omega_2+\omega_4)\Upsilon(\omega_1, -\omega_3)\Upsilon(\omega_2, -\omega_4) 
\end{eqnarray}
\end{small}
where 
\begin{eqnarray}
\Gamma(\omega, \omega')&\equiv& A(\omega)B(-\omega')+C(\omega)D(-\omega') \\
\Upsilon(\omega, \omega')&\equiv& B(\omega)B(-\omega')+D(\omega)D(-\omega'). 
\end{eqnarray}
Performing one integral for each delta function, we arrive to an expression that is $t$-independent
\begin{eqnarray}
\label{step2}
G^{(2)}(T)&=&\{\mathcal{F}[\Gamma](T)\}^2 +\{\mathcal{F}[\Upsilon](T)\}^2+\{\mathcal{F}[\Upsilon](0)\}^2 
\end{eqnarray}
where $\Gamma(\omega) \equiv\Gamma(\omega, \omega)$ and $ \Upsilon(\omega)\equiv\Upsilon(\omega, \omega)$.
Knowing that $r_1^2+t_1^2=1$ and $r_2^2+t_2^2=1$, from Eqs. (\ref{eq:Adef})--(\ref{eq:Ddef}) we find
\begin{eqnarray}
\label{eq:GammaDef}
\Gamma(\omega)&=& d(\omega)d(-\omega) t_1^2 \sinh(r)  \left[(1+r_1^2 r_2^2)\cosh(r) 
\right. \nonumber \\ & & \left. - r_1 r_2 e^{i\omega\tau}- r_1 r_2 e^{-i\omega \tau}\right], \\
\label{eq:UpsilonDef}
\Upsilon(\omega)&=& d(\omega)d(-\omega) t_1^2 \sinh(r)^2 (1-r_1^2 r_2^2).  
\end{eqnarray}
The necessary Fourier transforms are computed in the Appendix, see Eqs. (\ref{gammaF}) and (\ref{upsilonF}), in terms of a function $F(k)$, defined in Eq. (\ref{eq:FDef}).  We find  
\begin{eqnarray}
\label{eq:GammaFourier}
\{\mathcal{F}[\Gamma](T)\}^2&=&t_1^4 \sinh(r)^2  \sum_{k=-\infty}^\infty \delta(T-k\tau)
 \\ 
& & \times \left[(1+r_1^2 r_2^2)\cosh(r) F(|k|)  \right.
\nonumber \\ & & - r_1 r_2F(|k|+1)
 \left. - r_1 r_2 F(|k|-1) \right]^2 \nonumber \\ 
 \label{eq:UpsilonFourier}
\{\mathcal{F}[\Upsilon](T)\}^2&=&t_1^4 \sinh(r)^4 (1-r_1^2 r_2^2)^2  \\
& & \times \sum_{k=-\infty}^\infty \delta(T-k\tau)F(|k|)^2 \nonumber \\
\{\mathcal{F}[\Upsilon](0)\}^2&=&t_1^4 \sinh(r)^4 (1-r_1^2 r_2^2)^2 F(0)^2,
\end{eqnarray}
 the three terms necessary to calculate  $G^{(2)}(T)$. 
As shown in  Fig. \ref{theoryplot}, $G^{(2)}(T)$ of the \multimode cavity output has an envelope similar to the shape of double falling exponential and peaks  every cavity roundtrip time, resulting from the interference between the modes. In contrast, the single mode  $G^{(2)}(T)$ would also have a double exponential decay, but without the comb structure \cite{lu2000}.

\section{Comparison with earlier work}
The $G^{(2)}(T)$ calculation of  Lu and Ou \cite{lu2000} found the \multimode $G^{(2)}(T)$ to be a comb of (approximate) Dirac delta functions spaced by the cavity round-trip time, multiplied by an envelope given by the single-mode $G^{(2)}(T)$. 
This result has an appealing simplicity, and is intuitive in the time-domain picture in which photon pairs are produced simultaneously but may spend a different number of round trips in the cavity before escaping.  It is interesting to ask whether the same behaviour persists also at higher gains, i.e. in the presence of stimulated SPDC.  

We compare our $G^{(2)}(T)$, Eq. (\ref{step2}), against the natural extension of the Lu and Ou model for arbitrary gain, but still within the high-finesse approximation.  In this section we follow the notation of Refs. \cite{lu2000} and \cite{collett1984},
 and write $\exp[{-\gamma_i\tau}]=r_i$ to describe losses and $2\epsilon = r$ to describe gain. The single mode Bogoliubov transformations from \cite{collett1984}, without the low-gain approximation, are 
 \begin{figure}
\newcommand{\smallfigwidth}{.2 \textwidth}
  \centering
  \begin{tabular}{cc}
    \includegraphics[width=6.5cm]{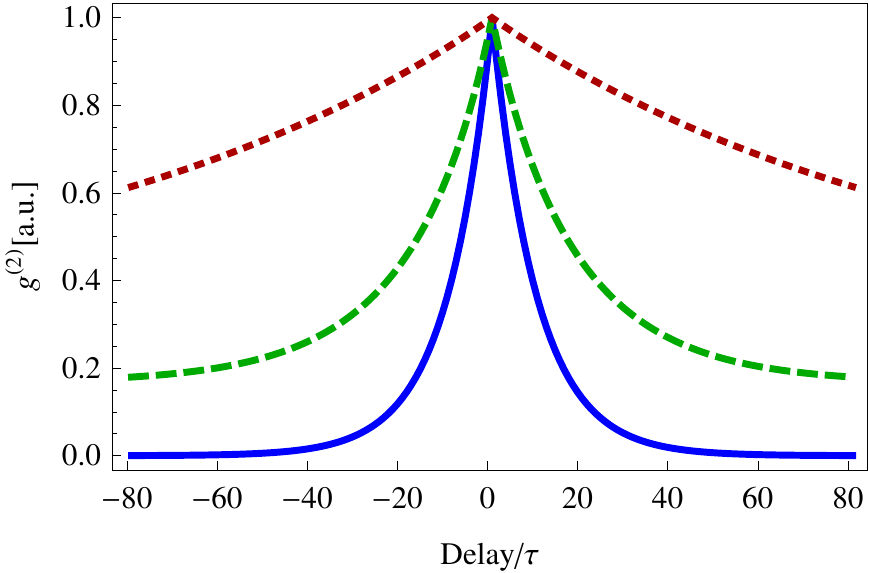}&
  \end{tabular}
  \caption{(color online) Three envelopes of \multimode $G^{(2)}(T)$, computed from  Eqs. (\ref{step2}), (\ref{eq:GammaFourier}) and (\ref{eq:UpsilonFourier}) and normalized to unity at $T=0$.  Curves show $G^{(2)}(T)$ for the gain $r$ equal to 1$\%$ (blue), 50$\%$ (green, dashed), 90$\%$ (red, dotted) of the threshold gain $r_{th}$.  Cavity parameters are as for the source presented in \cite{Anacavity}.}
  \label{pumppowers}
\end{figure}
\begin{eqnarray}
\label{eq:Asdef}
A_{\rm single}(\omega)&\equiv&  \frac{(\gamma_1/2)^2 - (\gamma_2/2 - i \omega)^2 + |\epsilon|^2}{(\gamma_1/2 + \gamma_2/2 -i \omega)^2 - |\epsilon|^2}\\
B_{\rm single}(\omega)&\equiv& \frac{\gamma_1 \epsilon}{(\gamma_1/2 + \gamma_2/2 - i \omega)^2 - |\epsilon|^2}\\
C_{\rm single}(\omega)&\equiv& \frac{\sqrt{\gamma_1 \gamma_2} (\gamma_1/2 + \gamma_2/2 - i \omega)}{(\gamma_1/2 + \gamma_2/2 - i \omega)^2 - |\epsilon|^2}\\  
D_{\rm single}(\omega)&\equiv& \frac{\sqrt{\gamma_1 \gamma_2} \epsilon}{(\gamma_1/2 + \gamma_2/2 - i \omega)^2 - |\epsilon|^2} .
\end{eqnarray}
We  follow the same steps as from Eq. (\ref{step1}) to Eq. (\ref{step2}),  
to find
\begin{eqnarray}
\label{step2single}
G^{(2)}_{\rm single}(T)&=&\{\mathcal{F}_{\rm single}[\Gamma](T)\}^2 +\{\mathcal{F}_{\rm single}[\Upsilon](T)\}^2
\nonumber \\
& & +\{\mathcal{F}_{\rm single}[\Upsilon](0)\}^2 
\end{eqnarray}

where 
\begin{eqnarray}
\{\mathcal{F}_{\rm single}[\Gamma](T)\}^2&=& \frac{\pi}{2} \gamma_1^2 \epsilon^2 \left( f_-  + f_+  \right)^2\\\nonumber\\
\{\mathcal{F}_{\rm single}[\Upsilon](T)\}^2&=& \frac{\pi}{2} \gamma_1^2 \epsilon^2 \left( f_-  - f_+  \right)^2\\\nonumber\\
f_{\pm} & \equiv & \frac{e^{-\frac{1}{2}|T|(\gamma_1 + \gamma_2 \pm 2 \epsilon)}}{\gamma_1 + \gamma_2 \pm 2 \epsilon}
\end{eqnarray}

Finally, we multiply by a comb of (approximate) delta functions. Again following \cite{lu2000}, for a \multimode cavity with $2N+1$ modes we have:

\begin{eqnarray}
\label{eq:GmultiSin}
   G_{\rm multi}^{(2)}(T) &\propto & G_{\rm single}^{(2)}(T) \frac{\sin^2[(2N+1)\pi T/ \tau]}{\sin^2[\pi T/\tau]}
 \\ 
\lim_{N\rightarrow\infty} G_{\rm multi}^{(2)}(T) &\propto &  G_{\rm single}^{(2)}(T)\sum_{n=-\infty}^{\infty}\delta(T-n\tau).
\end{eqnarray}

Eq. (\ref{step2single}), computed by extension of \cite{lu2000}, agrees very closely with our \multimode result Eq. (\ref{step2}), shown in Fig. \ref{pumppowers}.  The only situation for which the two approaches give significantly different results is when the output coupler has high transmission $t_1$. Even so, the difference between the two calculations  does not exceed $7.5\%$ of the value of $G^{(2)}(T)$, for $r_1,r_2>0.5$ and for any sub-threshold gain.  We conclude that for many purposes the very simple results of Eq. (\ref{step2single}) can be used, backed by the more accurate calculation given in Section \ref{sec:MMG2}.

Fig. \ref{pumppowers} shows the computed shape of the $G^{(2)}(T)$ envelope as a function of gain parameter $r$.  This clearly shows a broadening of the correlations, along with a raising of the background level, which persists to arbitrarily large $|T|$.  The background can be understood as a result of ``accidental'' coincidences, i.e. correlations among photons that were not produced in the same SPDC event.  The broadening is the time-domain manifestation of the narrowing of the resonances with increasing $r$, visible e.g. in $d(\omega)$.  Physically, it can be understood as the coherent amplification of SPDC photons already inside the cavity, i.e., stimulated SPDC.  This change in photon temporal distributions is of potential interest in wave-function matching for non-classical interference \cite{PatelNJP2014}, matching to quantum memories \cite{SpragueNPhot2014},
 and detection of ``Schr{\"o}dinger kittens" and other time-localized non-classical fields \cite{MorinPRL2013}.

\section{Conclusion}
We have computed the output of a \multimode cavity-enhanced spontaneous parametric down-conversion source, including realistic mode structure and sub-threshold but otherwise arbitrary gain. Using time-domain difference equations describing field operators at  consecutive roundtrips, we find \multimode Bogoliubov transformations that describe the output field.  This analytic solution provides a basis for calculations of any correlation function describing the \multimode output.  We compute the two-time intensity correlation function $G^{(2)}(T)$, and find increased temporal coherence due to stimulated SPDC in both single and \multimode cases.  We extend a calculation by Lu and Ou \cite{lu2000} to arbitrary gain, and find that it agrees well with our more exact calculation.  The results will be useful in describing high-gain spontaneous parametric down-conversion, in the context of quantum networking using atomic quantum memories \cite{PomaricoNJP2009,ChuuAPL2012, PomaricoNJP2012, FeketePRL2013} and studies of ``Schr{\"o}dinger kittens'' and other exotic non-classical states  \cite{NeergaardNielsenPRL2006,AndersenPRA2013,MorinNPhot2014,JeongNPhot2014}.

\section{Acknowledgements}
This work was supported by the Spanish MINECO project MAGO (Ref.
FIS2011-23520) and the European Research Council project AQUMET, and Fundaci\'{o} Privada CELLEX.   J. A. Z. was supported by the FI-DGR PhD-fellowship program of the Generalitat of Catalonia.
\begin{appendix}
\section{Fourier transforms for $\Gamma$ and $\Upsilon$}

We first compute ${\cal F}[d(\omega)d(-\omega)](T)$, the Fourier transform  of $d(\omega) d(-\omega)$, where $d$ is given in Eq. (\ref{eq:denominator}).  We denote  $x \equiv ({1 + r_1^2 r_2^2e^{2r}})/({2 r_1 r_2 e^r} )$ and $y \equiv ({1 + r_1^2 r_2^2e^{-2r}})/({2 r_1 r_2 e^{-r}})$. In the below-threshold regime we are considering, $r<r_{\rm th}=-\log(r_1 r_2)$ so that $d(\omega)$ is always finite.  We find
\begin{eqnarray}
d(\omega)d(-\omega)=\frac{1}{4 r_1^2 r_2^2}\frac{1}{x-\cos(\omega\tau)}\frac{1}{y-\cos(\omega\tau)}
\end{eqnarray}
Since $d(\omega)d(-\omega)$ is an even periodic function with a period of $2 \pi/\tau$ we can write
\begin{eqnarray}
d(\omega)d(-\omega)=\sum_{k=0}^\infty F(k) \cos(k\omega\tau)
\end{eqnarray}
Where
\begin{eqnarray}
F(k)=\frac{2}{\pi}\int_0^\pi d(\omega)d(-\omega)\cos(k\omega\tau)d\omega
\end{eqnarray}
The Fourier transform is then the sum of  Dirac delta functions:
\begin{small}
\begin{eqnarray}
\label{delta}
\mathcal{F}[d(\omega)d(-\omega)](T)=\sum_{k=-\infty}^\infty F(|k|) \delta(T-k\tau)
\end{eqnarray}
\end{small}

The $F(k)$ can be expressed in terms of hypergeometric functions
\begin{eqnarray}
\label{eq:FDef}
F(k)&=&\frac{2}{4 r_1^2 r_2^2}\frac{1}{(x-y)(1+x)(1+y)}
 \\ &  & \times
 \left[(1+x)\frac{{}_3F_2 \left(\{\frac{1}{2},1,1\},\{1-k,1+k\};\frac{2}{1+y} \right)}{\Gamma(1-k)\Gamma(1+k)} \right. \nonumber \\ & & \left.  -(1+y)\frac{{}_3F_2 \left(\{\frac{1}{2},1,1\},\{1-k,1+k\};\frac{2}{1+x} \right)}{\Gamma(1-k)\Gamma(1+k)}\right]. \nonumber 
\end{eqnarray}
It follows immediately that the Fourier transform of 
$d(\omega)d(-\omega) e^{in\omega\tau}$ is 
\begin{eqnarray}
\label{eq:DDOmFour}
{\cal F}[d(\omega)d(-\omega) e^{in\omega\tau}](T) & = &  \sum_{k=-\infty}^\infty F(|k|+n)\delta(T-k\tau). \nonumber \\
\end{eqnarray}
Now in order to compute $\{\mathcal{F}[\Gamma](T)\}^2$ and $\{\mathcal{F}[\Upsilon](T)\}^2$, let us use the following trick. For a moment, let's assume that the bandwidth of the downconversion is finite, i.e. replace squeezing amplitude $r$ by a function $ r  \rm{rect} (\omega/\omega_{\rm bw})$ where
\begin{equation}
\rm{rect} (x)=\begin{cases}
    1,& \text{if } |x|<1/2\\
    0,              & \text{otherwise}
\end{cases}
\end{equation}
later we will apply to the final expressions the limit $\omega_{\rm bw}\rightarrow\infty$ returning to the situation with the infinite bandwidth.
In that case the functions $\Gamma_{\rm bw}(\omega)$ and $\Upsilon_{\rm bw}(\omega)$ yield
\begin{eqnarray}
\Gamma_{\rm bw}(\omega)= \rm{rect} (\omega/\omega_{\rm bw}) \Gamma(\omega)
\end{eqnarray}
\begin{eqnarray}
\Upsilon_{\rm bw}(\omega)= \rm{rect} (\omega/\omega_{\rm bw}) \Upsilon(\omega)
\end{eqnarray}
Therefore, if we write $*$ for convolution we find
\begin{eqnarray}
\{\mathcal{F}[\Upsilon_{\rm bw}](T)\}= \frac{ \omega_{\rm bw}^2}{\sqrt{2\pi}}\{\mathcal{F}[\Upsilon](T)\}*\rm{sinc}\left(\frac{T \omega_{\rm bw}}{2\pi}\right)
\end{eqnarray}
Knowing that
\begin{eqnarray}
\{\mathcal{F}[\Upsilon](T)\}= t_1^2 \sinh(r)^2 (1-r_1^2 r_2^2)\sum_{k=-\infty}^\infty \delta(T-k\tau)F(|k|) \nonumber \\
\end{eqnarray}
we arrive to 
\begin{eqnarray}
\{\mathcal{F}[\Upsilon_{\rm bw}](T)\}^2&=&t_1^4 \sinh(r)^4 (1-r_1^2 r_2^2)^2  \\
& & \times \left[\sum_{k=-\infty}^\infty \rm{sinc}\left(\frac{(T-k\tau) \omega_{\rm bw}}{2\pi}\right)F(|k|)\right]^2. \nonumber \\
\end{eqnarray}  
Now let's notice that for $k\neq l$
\begin{equation}
\lim_{\omega_{\rm bw}\rightarrow\infty}\rm{sinc}\left(\frac{(T-k\tau) \omega_{\rm bw}}{2\pi}\right)\rm{sinc}\left(\frac{(T-l\tau) \omega_{\rm bw}}{2\pi}\right)=0
\end{equation}
and
\begin{equation}
\lim_{\omega_{\rm bw}\rightarrow\infty}\left[\rm{sinc}\left(\frac{T \omega_{\rm bw}}{2\pi}\right)\right]^2=\delta(T)
\end{equation}
in the sense of a weak limit, i.e.
\begin{equation}
\lim_{\omega_{\rm bw}\rightarrow\infty}\int_\infty^\infty dT f(T)\left[\rm{sinc}\left(\frac{T \omega_{\rm bw}}{2\pi}\right)\right]^2=f(0)
\end{equation}
for any continuous function $f$ with a compact support.
It follows that:
\begin{eqnarray}
\label{upsilonF}
\{\mathcal{F}[\Upsilon](T)\}^2&=&t_1^4 \sinh(r)^4 (1-r_1^2 r_2^2)^2  \\
& & \times \sum_{k=-\infty}^\infty \delta(T-k\tau)F(|k|)^2 
\end{eqnarray}
An analogous argument leads to 
\begin{eqnarray}
\label{gammaF}
\{\mathcal{F}[\Gamma](T)\}^2&=&t_1^4 \sinh(r)^2  \sum_{k=-\infty}^\infty \delta(T-k\tau)
 \\ 
& & \times \left[(1+r_1^2 r_2^2)\cosh(r) F(|k|)  \right.
\nonumber \\ & & - r_1 r_2F(|k|+1)
 \left. - r_1 r_2 F(|k|-1) \right]^2. \nonumber 
\end{eqnarray}

\end{appendix}

\bibliography{FDC_bib}{}

\end{document}